\documentclass[twocolumn,showpacs,preprintnumbers,amsmath,amssymb,pre]{revtex4}

\usepackage{times,graphicx,amssymb,amsmath,psfig,epsfig}%,doublespace}

\begin{document}

\title{Emergence of time-horizon invariant correlation structure in financial returns by subtraction of the market mode}

\author{Christian Borghesi}
\affiliation{Service de Physique de l'Etat Condens\'e (CNRS URA 2464), CEA Saclay,
91191 Gif sur Yvette Cedex, France}

\author{Matteo Marsili}
\email{marsili@ictp.it}
\affiliation{The Abdus Salam ICTP, Strada Costiera 11, I-34014, Trieste. Italy.}\altaffiliation[also ]{DEMOCRITOS}

\author{Salvatore Miccich\`e}
\email{micciche@lagash.dft.unipa.it}
\affiliation{Universit\`a degli Studi di Palermo, Dipartimento di Fisica e Tecnologie Relative, Viale delle Scienze, Edificio 18, I-90128 Palermo, Italy.}

\begin{abstract}
  We investigate the emergence of a structure in the correlation
  matrix of assets' returns as the time-horizon over which returns are
  computed increases from the minutes to the daily scale. We analyze
  data from different stock markets (New York, Paris, London, Milano) and with
  different methods. Result crucially depends on whether the data is
  restricted to the ``internal'' dynamics of the market, where the
  ``center of mass'' motion (the market mode) is removed or not. If
  the market mode is not removed, we find that the structure emerges,
  as the time-horizon increases, from splitting a single large cluster.
  In NYSE we find that when the market mode is removed, the structure of
  correlation at the daily scale is already well defined at the 5
  minutes time-horizon, and this structure accounts for 80 \% of the classification of
  stocks in economic sectors. Similar results, though less sharp, are found for the
  other markets. We also find that the structure of correlations in the overnight 
  returns is markedly different from that of intraday activity. 
\end{abstract}

%% 89.75.Hc Networks and genealogical trees
%% 05.90.+m Other topics in statistical physics...
\date{\today}

\maketitle

%%%%%%%%%%%%%%%%%%%%%%%%%%%%%%%%%%%%%%%%%%%%%%%%%%%%%%%%%%%%%%%%%%%%%%%%%%%%%%%%%%%%%%%%%%%
\section{Introduction}
%%%%%%%%%%%%%%%%%%%%%%%%%%%%%%%%%%%%%%%%%%%%%%%%%%%%%%%%%%%%%%%%%%%%%%%%%%%%%%%%%%%%%%%%%%%

Besides their intrinsic interest, financial markets have also
attracted a great deal of attention as a paradigm of complex systems
of interacting agents. In this view, the correlation between different
assets are one of the signatures of the complexity of the system's
interactions and, as such, have been the focus of intense recent
research \cite{JPB-book,MantegnaStanley}. The central object of study is the empirical covariance
matrix of a set of $N$ assets, whose elements are the Pearson's
correlation coefficients $C_{i,j}^{(\tau)}(T)$ between the log--returns of
assets $i$ and $j$ over a time-horizon $\tau$, measured on historical
time series of length $T$.  Early studies have focused mainly on daily
returns ($\tau=1$ day) and have shown that the bulk of the
eigenvalue distribution of the correlation matrix is dominated by
noise and described very well by random matrix theory \cite{jpb-rmt,Stanley}.
This ``noise'' band of noisy eigenvalues shrinks as $\sqrt{N/T}$ as
the length $T$ of dataset increases, but it is significant for
typical cases where $N$ and $T$ are of the order of some hundreds.
The few large eigenvalues which leak out of the noise background
contain significant information about market's structure.  The
taxonomy built with different methods
\cite{mantegna,Stanley,giada,states} from financial correlations
alone bears remarkable similarity with a classification in economic
sectors. This agrees with the expectation that companies engaged in
similar economic activities are affected by economic factors in a
similar way. With respect to their dynamical properties, it has been
found that, financial correlations are persistent over time
\cite{kertesz} and that they follow recurrent patterns \cite{states}.

Furthermore, correlations ``build up'' as the time-horizon $\tau$ on
which returns are measured increases, and they saturate for returns on
the scale of some days \cite{Drozdztau,jpb0507111}. This behavior,
known as the Epps effect\cite{epps}, is a manifestation of the process
of mutual information exchange across assets. It quantifies how this
information flow is ultimately ``incorporated'' into correlations, in
much the same way as information on single assets is incorporated into
their prices. Interestingly, it was found that such a process is much
faster today than in the past and more pronounced for more capitalized
stocks \cite{Drozdztau}.
It has also been remarked \cite{blm01,AdP} that the structure of
correlations changes as the time-horizon $\tau$ over which returns are
defined increases, i.e. that ``pictorially, the market appears as an
embryo which progressively forms and differentiates over time''
\cite{jpb0507111}. 

Here we shall take a closer look on the dependence of the structure of
correlations on the time-horizon $\tau$ and show that the observed
evolution of the market structure is due, to some extent, to the
dynamics of the market mode. Global correlations play a dominant role
at high frequency, thus giving rise to correlation structures which
are much more clustered than at the daily scale. However, if global
correlations are removed, the structure of correlations at the daily
scale, is largely preserved across time-horizons, down to a scale of 5
minutes for the most liquid market we have analyzed. Loosely speaking,
the network structure, after removing the market mode, appears fully 
formed and differentiated already at small
scales, it only grows in size (of correlations) as the time-horizon
increases. 

The effect of disentangling the effect of the market mode when computing pairwise
correlations between stocks is analogous to decomposing the dynamics of a complex interacting system in that of its center of mass and of its internal
coordinates. This is obvious in physics, where the center of mass dynamics is determined by external forces, whereas internal coordinates mainly respond to
inter-particle interaction forces. By analogy, our results suggest that in order to understand the dynamics of inter-asset correlations, it makes sense
to eliminate the effect of the ``center of mass''. 

The paper is organized as follows: in the next Section we discuss the datasets and how we build correlation matrices. Then we shall discuss the results of data clustering approach in Section \ref{dataclustering} first for NYSE and then for the other markets. The following Section deals with the Minimal Spanning Trees  approach. Finally we shall summarize our results and offer some concluding remarks.

%%%%%%%%%%%%%%%%%%%%%%%%%%%%%%%%%%%%%%%%%%%%%%%%%%%%%%%%%%%%%%%%%%%%%%%%%%%%%%%%%%%%%%%%%%%
\section{The data} \label{data}
%%%%%%%%%%%%%%%%%%%%%%%%%%%%%%%%%%%%%%%%%%%%%%%%%%%%%%%%%%%%%%%%%%%%%%%%%%%%%%%%%%%%%%%%%%%

In this paper we empirically investigate the ensemble behaviour of price returns for $4$ different markets: 
the New York Stock Exchange (NYSE), the London Stock
Exchange (LSE), the Paris Bourse (PB) and the Borsa Italiana (BI). 
All data refer to year 2002. 

The NYSE data are taken from the Trades and Quotes (TAQ) database maintained by NYSE \cite{NYSE}. In particular, 100 highly capitalized stocks were considered. For each stock and for each trading day we consider the time series of stock prices recorded transaction by transaction. Since transactions for different stocks do not happen simultaneously, we divide each trading day (lasting $6^h~30'$) into intervals of length $\tau$. For each trading day, we define $N_\tau$ intraday stock price proxies $p_i(t_k)$ of asset $i$, with $k=1, \cdots, N_\tau$. The proxy is defined as the transaction price detected nearest to the end of the interval (this is one possible way to deal with high-frequency financial data \cite{Dacorogna2001}). By using these proxies, we compute the price returns
\begin{equation}
                a_{i}^{(\tau)}(t) = \ln p_i (t) - \ln p_i (t - \tau) 
\end{equation}
at time-horizons $\tau$. The time-horizon used are $\tau=5,~15,~30,~65,~ 195$ minutes. For NYSE, values of $\tau$ are large enough that all the considered stocks have at least one transaction in each time interval.

The LSE data are taken from the ``Rebuild Order Book'' database, maintained by LSE \cite{LSE}. In particular, we consider only the electronic transactions for 92 highly traded stocks belonging to the SET1 segment of the LSE market. The trading activity has been defined in terms of the total number of transactions (electronic and manual) occurred in 2002. However, most of the transactions, a mean value of $75 \%$ for the 92 stocks, are of the electronic type. This market is commonly believed to be very active and can be regarded as a realization of a ``liquid'' market.
For each stock $i$ and for each trading day we consider the time series of stock price recorded transaction by transaction and generate $N_\tau$ intraday stock price proxies $p_i(t_k)$ according to the procedure explained above. For the LSE data, the time-horizon used were $5,~15,~30,~51,~ 102,~ 255$ minutes. Each trading day lasts $8^h~30'$.

The PB data are taken from the ``Historical Market Data'' database, maintained by EURONEXT \cite{EURONEXT}. In particular, we consider the electronic transactions of two subsets of stocks traded in the year 2002. For each stock $i$ and for each trading day, lasting $8^h~30'$, we consider the time series of stock price recorded transaction by transaction and generate $N_\tau$ intraday stock price proxies $p_i(t_k)$ according to the procedure explained above. One first set, which will be analyzed in Section \ref{dataclustering}, consists of the $75$ most frequently traded stocks at time-horizons $\tau_k=27\cdot 2^k$ seconds, for $k=0,\ldots,10$. An analogous dataset was derived considering tick time: $\tau_k^{(tick)}=100\cdot 2^k$. This choice was considered in order to probe the region of very high frequencies and to assess the relevance of time inhomogeneity of trading activity at intraday time scales. In this respect, it is worth to remark that for small $\tau$ stocks were not traded in each time interval. 
A second dataset, that will be considered in Section \ref{SLCA}, instead consisted of $N= 39$ stocks which were continuosly traded in the entire 2002 (i.e. in each time interval) over  time-horizon of $\tau=5,~15,~30,~51,~ 102,~ 255$ minutes. 

The BI data are taken from the ``Dati Intraday'' database, maintained by Borsa Italiana \cite{BI}. In particular, we consider only the electronic transactions occurred for 30 stocks continuosly traded in the entire year 2002. For each stock $i$ and for each trading day we consider the time series of stock price recorded transaction by transaction and generate $N_\tau$ intraday stock price proxies $p_i(t_k)$ according to the procedure explained above. For the BI data, the time-horizon used were $5,~15,~30,~60,~ 120,~ 240$ minutes. Each trading day lasts $8^h$.

For all markets, in addition to the intraday time-horizons, we have considered returns on the daily time-horizon
\begin{eqnarray}
       a_{i}^{({\rm op-cl})}(n)&=&\log p_i^{\rm cl}(n)-\log p_i^{\rm op}(n),\\
       a_{i}^{({\rm cl-cl})}(n)&=&\log p_i^{\rm cl}(n)-\log p_i^{\rm cl}(n-1), \\
       a_{i}^{({\rm night})}(n)&=&\log p_i^{\rm op}(n)-\log p_i^{\rm cl}(n-1),
\end{eqnarray}
corresponding to intraday, daily and overnight returns, respectively. 
Here $p^{\rm op}_i(n)$ and  $p^{\rm cl}_i(n)$ are the open and closure prices of stock $i$ in day $n$.

%In order to check whether
%the structure of correlations is dominated by the initial and final
%period of transactions, we considered also reduced datasets where the
%initial and final period of $30$ ($20$) minutes for NYSE (PB) was
%removed.

Each stock can be associated to an economic sector of activity. For the NYSE data we considered the classification scheme given in the web--site {\tt{http://finance.yahoo.com/}}, for the LSE and BI data we considered the classification scheme used in the web--site {\tt{www.euroland.com}}, for the PB data we considered the classification scheme used in the web--site {\tt{http://www.euronext.com/}}. The relevant economic sectors are reported in Table \ref{classification}.

\begin{table}
\begin{center}
\caption{Color codes for the economic sectors of activity for the stocks.} \label{classification}
\vspace{.5 cm}
\begin{tabular}{||c|l|l||}
\hline
                     & ${\rm{SECTOR}}$      ~&~{\rm{COLOR}}~      \cr \hline
                  1  & Technology            & {\rm{red}}~        \\
                  2  & Financial             & {\rm{green}}~      \\
                  3  & Energy                & {\rm{blue }}~      \\
                  4  & Consumer non-Cyclical & {\rm{yellow}}~     \\
                  5  & Consumer Cyclical     & {\rm{brown}}~      \\
                  6  & Healthcare            & {\rm{grey}}~       \\
                  7  & Basic Materials       & {\rm{violet}}~     \\
                  8  & Services              & {\rm{cyan}}~       \\
                  9  & Utilities             & {\rm{magenta}}~    \\
                 10  & Capital Goods         & {\rm{light green}} \\
                 11  & Transportation        & {\rm{maroon}}~     \\        
                 12  & Conglomerates         & {\rm{orange}}~     \cr \hline
\end{tabular}
\end{center}
\end{table} 

Given the price return at a selected time-horizon $\tau$, we built the correlation matrices in the usual way
\begin{equation}
A_{i,j}^{(\tau)}=\frac{\langle a_i^{(\tau)}a_j^{(\tau)} \rangle
  -                    \langle a_i^{(\tau)} \rangle
                       \langle a_j^{(\tau)} \rangle}
                      {\sqrt{\langle [a_i^{(\tau)} - 
                                      \langle a_i^{(\tau)}\rangle
                                     ]^2 \rangle
                             \langle [a_j^{(\tau)} - 
                                      \langle a_j^{(\tau)}\rangle
                                     ]^2 \rangle}}.
\label{eq:Aij}
\end{equation}
Here and in what follows,
$\langle \ldots \rangle=(1/N_\tau)\sum_{t=1}^{N_\tau}\ldots$ denotes time
average.

In order to disentangle different components of the dynamics and to
understand their effect, we considered also series of datasets
derived from $a_i^{(\tau)}(t)$. In all derived datasets we subtract
a particular component of market dynamics from the rest. When
the structure of the derived dataset differ substancially from that
of the matrix $\hat A$ we can conclude that the decomposition is
meaningful and informative.

First we removed the ``center of mass'' dynamics:
\begin{equation}
                b_{i}^{(\tau)}(t)=a_i^{(\tau)}(t)-\frac{1}{N}\sum_{j=1}^N a_j^{(\tau)}(t).
\end{equation}
From this, a covariance matrix $B_{i,j}^{(\tau)}$ was computed in the
same way as in Eq. (\ref{eq:Aij}).

In a further dataset we removed the effect of the market index from
$a_i^{(\tau)}(t)$. This was done first considering the time-series
$I^{(\tau)}(t)$ of the corresponding market index at the same time-horizon $\tau$ and then estimating the coefficients of a one factor model
\begin{equation}
                a_{i}^{(\tau)}(t)=\alpha_i+\beta_i I^{(\tau)}(t) + c_{i}^{(\tau)}(t). \label{factor}
\end{equation}
The residuals $c_{i}^{(\tau)}(t)$ were used to build the 
covariance matrix $C_{i,j}^{(\tau)}$. We could build the time series $I^{(\tau)}(t)$
only in the case of NYSE data, for which we had access to intraday data of the SP500 composite index.

In all datasets we computed an ``endogenous'' market index using the market average return
\[
\bar{a}^{(\tau)}(t)=\frac{1}{N}\sum_{j=1}^N a_j^{(\tau)}(t).
\]
Using this instead of the market index $I^{(\tau)}(t)$ in Eq. (\ref{factor}) and considering the residues $d_{i}^{(\tau)}(t)$, we computed a further covariance matrix $D_{i,j}^{(\tau)}$.

Finally, we produced a dataset by removing the contribution of the
largest eigenvector of the matrix $A_{i,j}^{(\tau)}$. This can be done
by zeroing the largest eigenvalue of $\hat A$, as discussed in Ref.
\cite{Stanley}. An alternative method, which we prefer, is that of
removing the ``optimal'' factor, $G^{(\tau)}(t)$ which is obtained by
minimizing
\[
\chi^2=\sum_{i=1}^N\sum_{t=1}^{N_\tau}\left[
a_i^{(\tau)}(t)-\alpha_i-\beta_i G^{(\tau)}(t) \right]^2 \label{endo}
\]
on $\alpha_i$, $\beta_i$ and $G^{(\tau)}(t)$. The residues
$e_i^{(\tau)}(t)$ resulting from this operation coincide with the
time-series obtained from $a_i^{(\tau)}(t)$ by subtracting the leading
contribution of its singular value decomposition. We call
$E_{i,j}^{(\tau)}$ the correlation matrix of the residues $e_i^{(\tau)}(t)$.

In summary, we consider the original time-series (set $A$), the one
obtained subtracting the average market return (set $B$) and those
obtained from the residues of a one factor model with the market index
(set $C$), the average market return (set $D$) and the optimal factor
(set $E$). Set $C$ represents a case where the market mode is exogenously determined whereas in sets $D$ and $E$ it is determined by the data itself. This allows us to understand how much an index, such as SP500 which is a weighted average, accounts for the collective dynamics of the market.

%%%%%%%%%%%%%%%%%%%%%%%%%%%%%%%%%%%%%%%%%%%%%%%%%%%%%%%%%%%%%%%%%%%%%%%%%%%%%%%%%%%%%%%%%%%
%\subsection{The distribution of matrix elements}
%%%%%%%%%%%%%%%%%%%%%%%%%%%%%%%%%%%%%%%%%%%%%%%%%%%%%%%%%%%%%%%%%%%%%%%%%%%%%%%%%%%%%%%%%%%

\begin{figure}[ht]
%  \centerline{\hbox{\psfig{figure=hcij1.eps,height=6cm,width=8cm}}}
      \includegraphics[scale=0.4] {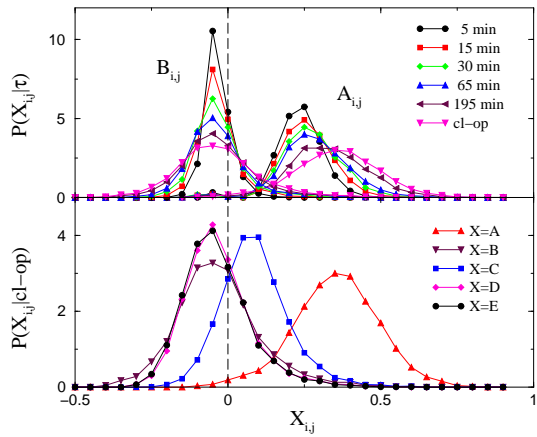} 
      \caption{Distribution of correlation coefficients $A_{i,j}$
      and $B_{i,j}$ for different time-horizons $\tau$ (top) and at
      the intraday time-horizon for different datasets (NYSE data).}
     \label{fig:hcij}
\end{figure}

The distribution of matrix elements is shown in Fig.
\ref{fig:hcij} as a function of time-horizon (top, for the sets $A$
and $B$) and for different datasets at the intraday time-horizon.
We observe that the distribution spreads out as the time-horizon
increases, as a manifestation of the Epps effect. However, while the distribution of $A_{i,j}$ is
centered around a positive value, that of correlations of derived
datasets is peaked on values close to zero and is narrower. For
set $B$ ($D$ and $E$) the peak is at slightly negative values,
whereas for set $C$ it occurs at positive values. This suggests
that the removal of correlations is more efficient when the single
factor is computed from the data. This already shows that the
dynamics of the mean $\bar{a}(t)$ already explains the
correlations better than the market index.

We also find that intraday and overnight returns have distinctly
different distribution of correlation coefficients. This
difference is particularly pronounced in dataset $C$ which again
suggest that the market index is even less explicative of the
market's collective behavior at these scales.

Correlation $D_{i,j}$ and $E_{i,j}$ were found to have a
distribution which is similar to that of $B_{i,j}$. This
anticipates a generic conclusion: the subtraction of a global
component from the dynamics is most meaningful when it eliminates
(either implicitly as in $\hat B$ or explicitly as in $\hat E$)
the market mode by setting the corresponding eigenvalue to zero.

\begin{figure}[ht]
%  \centerline{\hbox{\psfig{figure=eig.a.eps,height=6cm,width=8cm}}}
      \includegraphics[scale=0.6] {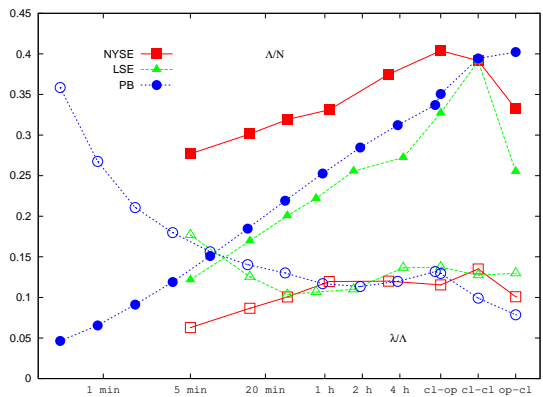} 
      \caption{Largest eigenvalue $\Lambda/N$, divided by the number of assets, of the matrix $\hat A_\tau$ as a function of $\tau$ for NYSE, LSE and PB (full symbols). Ratio $\lambda/\Lambda$ of the second largest to the largest eigenvalue of $\hat A_\tau$, as a function of $\tau$ (open symbols).}
     \label{fig:eig}
\end{figure}

Before analyzing the structure of correlations, it is of interest to provide some estimate of the relative strength of global correlations and of noise in the correlation matrices $\hat A$. Fig. \ref{fig:eig} plots the share of correlation carried by the largest eigenvalue $\Lambda$ (which is $\Lambda/N$, by normalization) for NYSE, LSE and PB, as a function of time-horizon $\tau$. 
As a manifestation of Epps effect \cite{epps}, this increases with $\tau$ in a way which is reasonably well approximated by a logarithmic growth. The ratio of the second largest eigenvalue $\lambda$ to the largest, which could be taken as a measure of the relative strength of inter-asset correlations against global correlations, has a declining trend with $\tau$ for small time-horizons and then saturates at around $0.1$. 

%%%%%%%%%%%%%%%%%%%%%%%%%%%%%%%%%%%%%%%%%%%%%%%%%%%%%%%%%%%%%%%%%%%%%%%%%%%%%%%%%%%%%%%%%%%
\section{Data Clustering}\label{dataclustering}
%%%%%%%%%%%%%%%%%%%%%%%%%%%%%%%%%%%%%%%%%%%%%%%%%%%%%%%%%%%%%%%%%%%%%%%%%%%%%%%%%%%%%%%%%%%

We performed data clustering analysis following the method of Ref.
\cite{giada}. Here we only sketch the basic idea of the method
and we refer the interested reader to Ref. \cite{giada} for
details. In brief, assume we wish to cluster $N$
standardized {\footnote{The time series $x_i(t)$ are derived from
                        $a_i^{(\tau)}(t),\ldots,e_i^{(\tau)}(t)$ 
                        by normalization to zero average and unit variance. E.g.
       $x_i(t)=[a_i^{(\tau)}(t)-\langle {a_i^{(\tau)}} \rangle ]/
       \sqrt{\langle [a_i^{(\tau)} -  \langle a_i^{(\tau)} \rangle ]^2 \rangle}
       $.}}
time series $x_i(t)$ in groups
having a similar dynamics. First we assign a cluster label $s_i$ to
each time series, specifying which cluster it belongs to. Then we
assume that $x_i(t)$ is generated according to the model
\begin{equation}
x_i(t)=g_{s_i}\eta_{s_i}(t)+\sqrt{1-g_{s_i}^2}\epsilon_i(t),~~~~t=1,\ldots,T
\label{model}
\end{equation}
where $\eta_s(t)$ and $\eta_i(t)$ are independent gaussian variables
with mean zero and unitary variance. Here $\eta_s(t)$ describes the
component of the dynamics which is common to all time series $x_i(t)$
with $s_i=s$ whereas $\epsilon_i(t)$ describes idiosyncratic
fluctuations. Eq. (\ref{model}) is consistent with a correlation
matrix $X_{i,j}=\langle x_ix_j \rangle$ which has a block diagonal structure for
$T\to\infty$: $X_{i,j}=g^2_{s_i}$ if $s_i=s_j$ and $X_{i,j}=0$
otherwise. The parameters $g_s$ entering Eq. (\ref{model}) as well as
the cluster structure $\{s_i\}$ can be determined by maximum
likelihood estimation. Approximate maximization of the log-likelihood
can be done following an hierarchical clustering procedure \footnote{We
  choose this simple option, rather than more elaborate maximization
  procedures based e.g. on simulated annealing, because for the data
  sets used the optimal configuration we find depend only marginally
  on the algorithm used.}: start with $N$ clusters, each composed of a
single asset ($s_i^{(0)}=i$). From the configuration $\{s_i^{(K+1)}\}$
with $K+1$ clusters, compute the log-likelihood of all configurations
obtained by merging two clusters. The configuration $\{s_i^{(K)}\}$
with $K$ clusters is the one corresponding to the maximal
log-likelihood ${\cal L}_K$. This operation can be iterated with $K$
going from $N-1$ to $1$, and the optimal configuration can be chosen
as that for which ${\cal L}_K$ is maximal. This also predicts the
optimal number $K^*$ of clusters which describes our dataset.
This method has already been used to analyze stock market data: in
Refs. \cite{giada} the emergent clusters were found
to be highly correlated with economic activity. Furthermore the method
was extended to perform noise undressing. In Ref. \cite{states} the
method has been applied to investigate market dynamics, showing that
well defined recurrent states of market wide activity can be defined.

Here we apply this method to investigate how the structure of market's
correlations evolves as the time lag $\tau$ increases from the
high-frequency range to the daily scale. We shall first focus on NYSE
and then discuss the differences found in other markets.

\begin{figure}[ht]
%  \centerline{\hbox{\psfig{figure=ncls.eps,height=6cm,width=8cm}}}
      \includegraphics[scale=0.4] {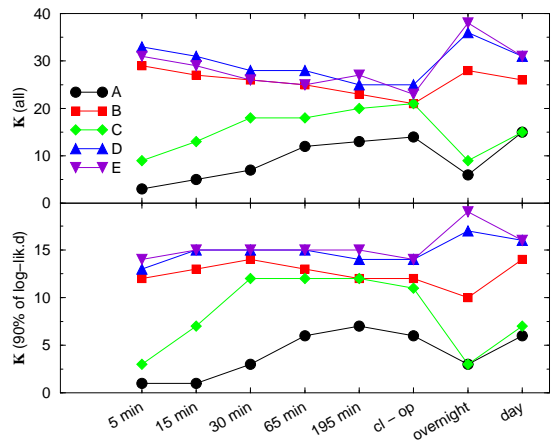} 
      \caption{Top: Number of clusters for datasets $A,B,C,D$ and
        $E$. Bottom: number of clusters accounting for 90\% of the
        likelihood (NYSE data).}
     \label{fig:ncls}
\end{figure}

\subsection{NYSE}

Fig. \ref{fig:ncls} shows the
evolution of the number of clusters with the time-horizon $\tau$ for the
different datasets in the NYSE. For $\hat A$ we find fewer clusters
then with other methods and the number of clusters increases with
$\tau$. This is consistent with results of Refs.
\cite{blm01} which observe an evolution of the structure of
correlations, where more and more details are added as the time-horizon
increases. The other datasets, however, reveal that this is due to
the fact that $\hat A$ includes the correlations induced by the common
factor. When this is removed, as for $\hat B, \hat D$ and $\hat E$, we
find that the number of clusters which accounts for most of the
log-likelihood is remarkably stable from the 5 min to the intraday
scale.  When the S\&P500 index is removed from the data ($\hat C$), we
find a fast evolution of the structure between 5 min and 30 min and then
the number of clusters saturates to a constant level.  Again, in all
cases, a significant variation takes place in the overnight and hence
at the daily (cl-cl) scale.

\begin{figure}[ht]
%  \centerline{\hbox{\psfig{figure=evol.cls.a.eps,height=6cm,width=8cm}}}
      \includegraphics[scale=0.32] {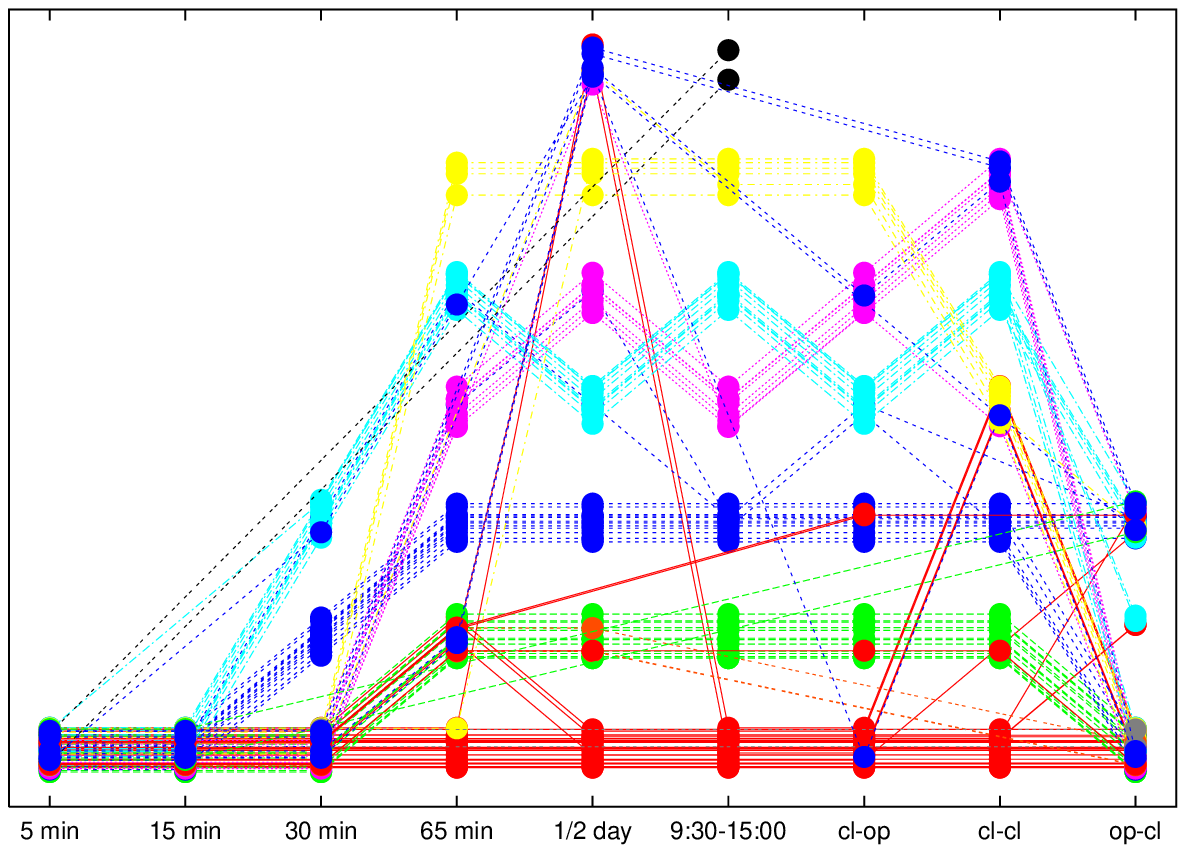} 
      \includegraphics[scale=0.32] {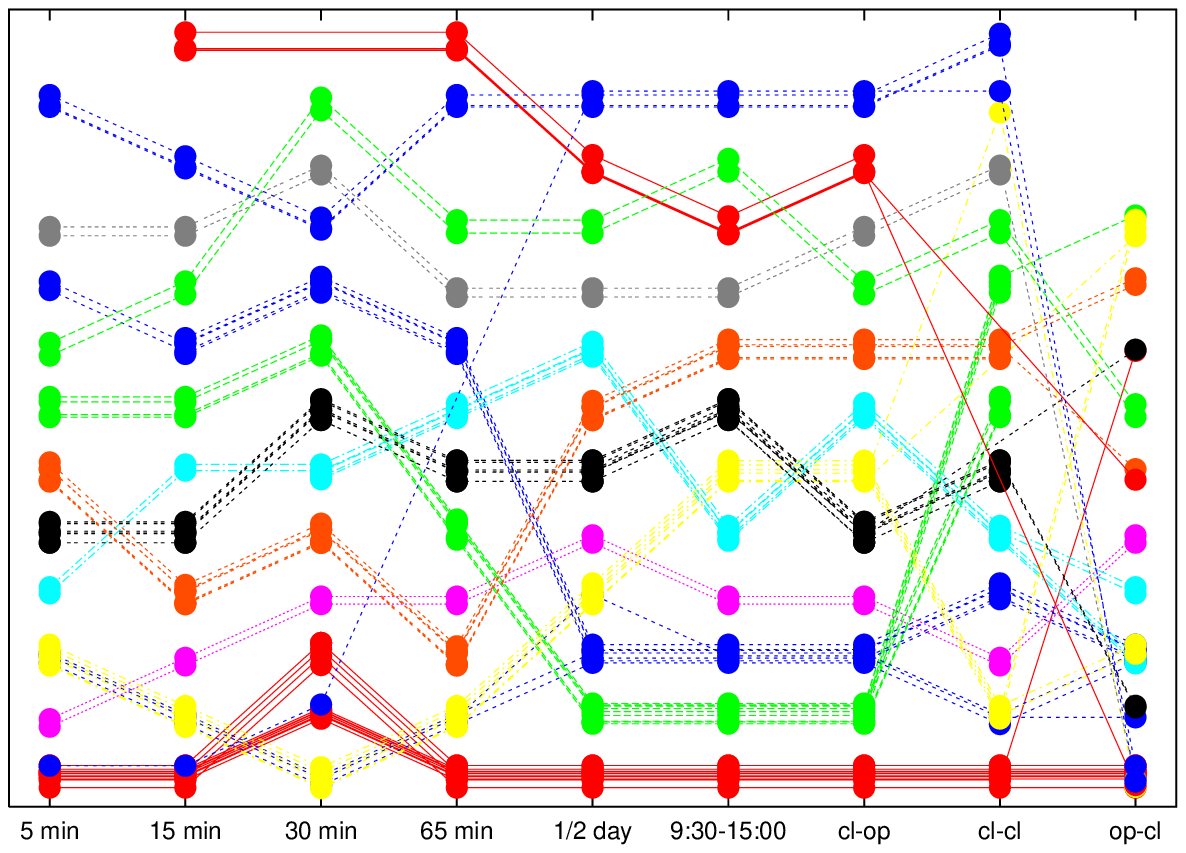} 
      \includegraphics[scale=0.32] {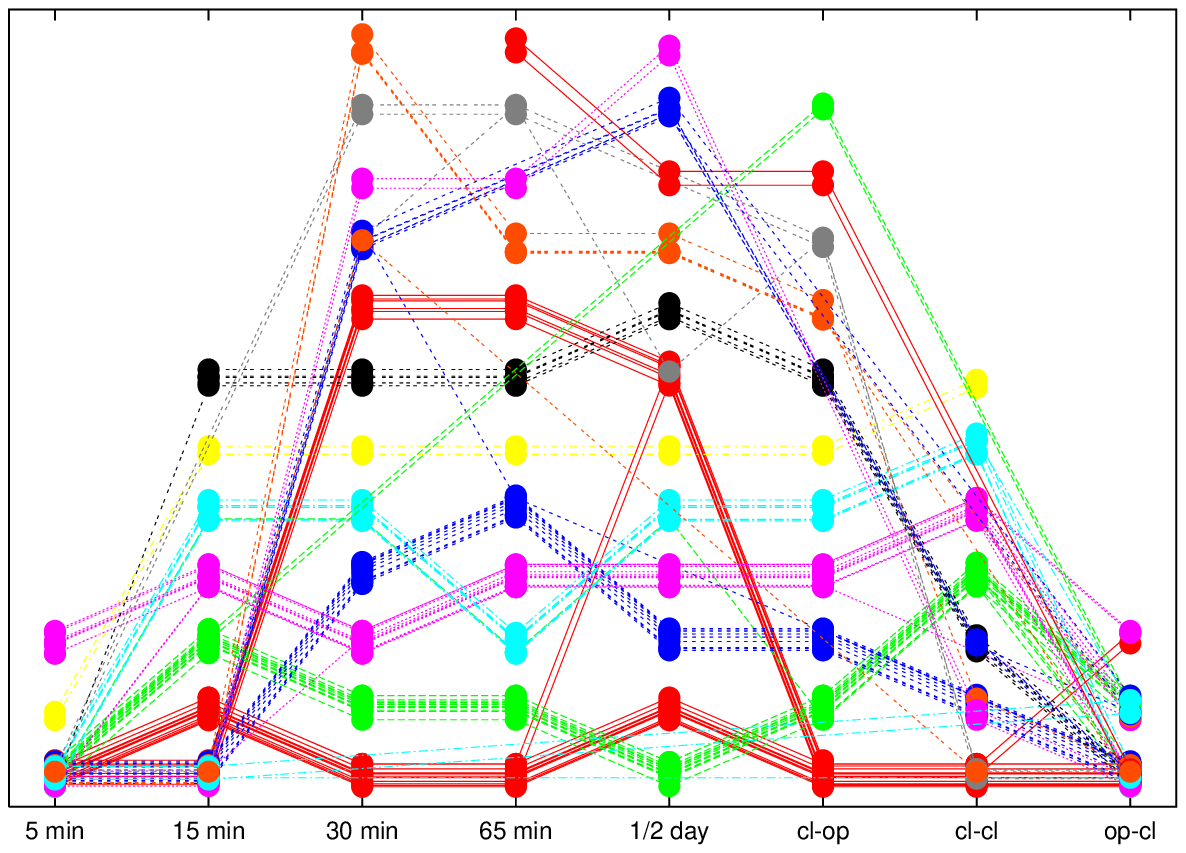} 
      \includegraphics[scale=0.32] {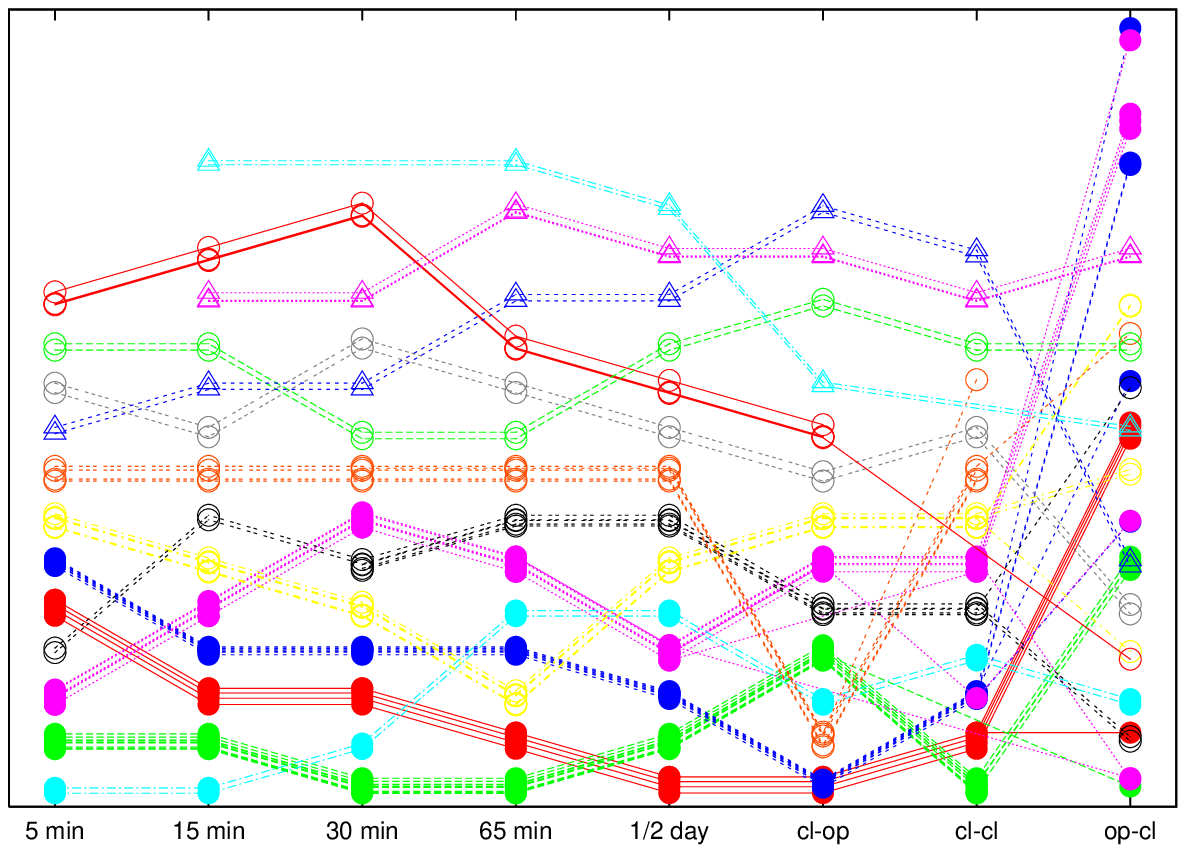} 
      \caption{Evolution of the cluster structure with time-horizon for the
      set $A$ (top left) $B$ (top right), $C$ (bottom left) and $E$ (bottom right) of NYSE. The cluster label $s_i^{(\tau)}$ of each asset belonging
      to the most relevant clusters is shown as a function of $\tau$. In
      this way, assets who always belong to the same cluster follow the
      same ``trajectory'' (indeed trajectories of different assets $i$ are
      shifted by a small random variable $\epsilon_i$ to distinguish them).
      The color is relative to the cluster structure at the intraday scale.}
     \label{fig:evol.cls}
\end{figure}

A closer view on the evolution of the cluster structure is presented
in Figs. \ref{fig:evol.cls}.  This plots the
cluster label $s_i^{(\tau)}$ as a function of $\tau$, for each asset
belonging to clusters accounting for 90\% of the
log-likelihood \footnote{Labels are sorted with respect to their
  contribution to the log-likelihood.  E.g. $s_i^{(\tau)}=3$ means
  that asset $i$ belongs to the third most relevant cluster at
  time-horizon $\tau$.}. Hence assets $i$ and $j$ belonging to the same
cluster for all $\tau$, follow parallel trajectories in the figure. In
this representation, cluster splitting and merging can clearly be read
off. In dataset $\hat A$ and $\hat C$ we see considerable splitting
of clusters as we move from $\tau=$5 min to the daily time-horizons. A
substantial reshuffling and merging takes place when going to
overnight returns. On the contrary, in dataset $\hat B$,  $\hat E$ and $\hat D$ (not shown), cluster membership exhibits a remarkable stability at intraday scales: the vast majority of assets within a cluster at 5 min, follows the same ``trajectory'' across 
time-horizons. Some reshuffling takes place in the order of clusters,
suggesting that the structure of correlations among sectors evolves
with time-horizons.  Again, the structure of overnight returns is
considerably different.

\begin{table}[htdp]
\begin{center}\begin{tabular}{|c|c|c|c|c|c|c|c|c|}\hline 
$\Im$(\%) & 5 min & 15 min & 30' & 65 min & 195 min & cl-op & cl-cl & op-cl \\\hline 
A &   5 &  11 &  42 &  77 &  86 & 100 &  89 &  24 \\
B &  91 &  90 &  91 &  90 &  92 &  90 &  90 &  72 \\ 
C &  33 &  66 &  84 &  86 &  87 &  92 &  89 &  30 \\ 
D &  91 &  90 &  92 &  91 &  89 &  92 &  90 &  78 \\ 
E &  91 &  87 &  90 &  87 &  90 &  90 &  90 &  80 \\ \hline
\end{tabular} \caption{Overlaps $\Im$ between cluster structures at different time-horizons and different sets and the structure of set $A$ at $\tau=1$ day.}
\end{center}
\label{tab:overlaps}
\end{table}

In order to make the comparison of different cluster structure quantitative, we have introduced an information distance $\Im(s^{(1)},s^{(2)})$ between any two structures $\{s^{(1)}_i\}$ and $\{s^{(2)}_i\}$. In words, this tells us how much the knowledge of the cluster label $s^{(1)}_i$ of a randomly chosen stock $i$, yields information on the value of $s_i^{(2)}$. Information is quantified by entropy reduction, in the following manner: Let $p^{(\ell)}(s)$ be the fraction of stocks with $s_i^{(\ell)}=s$ for $\ell=1,2$ and $p^{(1|2)}(s|s')$ be the fraction of stocks with $s_i^{(1)}=s$, among those which have $s_i^{(2)}=s'$. From these, we can compute the entropies $S^{(\ell)}$  in the usual way
and the conditional entropy
\[
S^{(1|2)}=-\sum_{s'}p^{(2)}(s')\sum_s p^{(1|2)}(s|s')\log p^{(1|2)}(s|s').
\]
The information gain is then given by
\begin{equation}
\Im = \frac{S^{(1)}-S^{(1|2)}}{S^{(1)}}
\end{equation}
Because of the normalization, a value of $\Im\approx 1$ implies that $s^{(2)}$ yields a rather precise information on $s^{(1)}$, so if $\Im = 0.8$ we shall say that $s^{(2)}$ accounts for 80 \% of the information contained in $s^{(1)}$.
Table \ref{tab:overlaps} shows the values of $\Im$ (in \%) between different cluster structures and that obtained from set $A$ at $\tau=$1 day time-horizon.
This shows that at this time-horizon, the cluster structure is essentially the same in the five datasets, with an overlap larger than 90 \%. An overlap of the same order of magnitude attains for all intraday scales in sets $B,D$ and $E$. Even though the overlap drops down as one moves to overnight returns, the difference is much smaller in sets  $B,D$ and $E$ than in sets $A$ and $C$. This suggests that, even though overnight returns have a structure which is markedly different from that of intraday returns, still removing the market mode allows one to reveal more invariant features.

\begin{figure}[ht]
%  \centerline{\hbox{\psfig{figure=relent.eps,height=6cm,width=8cm}}}
      \includegraphics[scale=0.6] {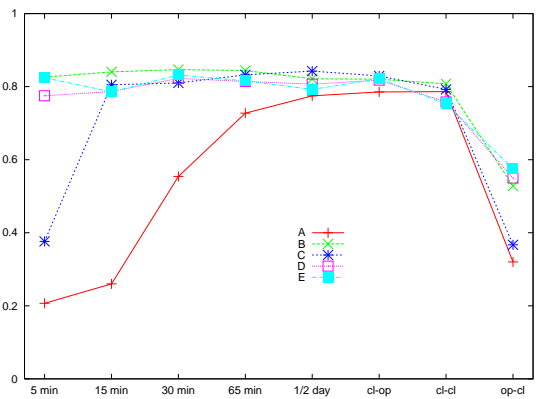} 
      \caption{Information gain on the classification in economic sectors given by the knowledge of cluster structures $s_i^{(\tau)}$ at different time-horizons $\tau$,  for different datasets $a,\ldots,e$. In order to avoid effects due to differences in the number of clusters, we considered maximum likelihood structures with $20$ clusters for all datasets. Notice that by normalization $0\le \Im \le 1$.}
     \label{fig:relent}
\end{figure}

Such invariant features, we claim, are related to economic sectors. In order to support this, we compare the cluster structures with the classification of assets in the sectors of economic activity given in Table \ref{classification}. The latter, yields a sector label $e_i\in \{1,\ldots,12\}$ for each stock $i$, for which we can compute an information gain $\Im$, as above, setting $s_i^{(1)}=e_i$. Fig. \ref{fig:relent} shows the behavior of  $\Im$ for different datasets across time-horizons. This suggests that the most informative sets are those where the market mode is removed and these account for 80\% of the information contained in $e_i$. For these, the information content is remarkably constant across time-horizons.
On the contrary, for set $A$ the information gain $\Im$ increases with $\tau$ in the intraday range, as if information on the economic activity of assets were "released" gradually, as time-horizon increases. It is worth to remark that, for all datasets, overnight returns (specially for sets $A$ and $C$) carry much less information on the economic structure of the market, than intraday returns.

Hence, we conclude that in datasets $A$ and $C$ the evolution in the
cluster structure is due to the interplay between the ``center of mass''
motion (i.e. the market mode) and the internal dynamics. Indeed when the
latter contribution is subtracted from the data, as in datasets $B$, $D$
and $E$, we find that the structure of correlations is remarkably stable
with the time-horizon. This is consistent with a notion of market's
informational efficiency by which information is incorporated very quickly
in market's returns. From the above analysis, we infer that the information
on the relations between assets is efficiently incorporated in returns
over time-horizons shorter than 5 min in NYSE.

\subsection{Other markets}

\begin{figure}[ht]
%  \centerline{\hbox{\psfig{figure=evol.cls.e.LSE.eps,height=6cm,width=8cm}}}
      \includegraphics[scale=0.6] {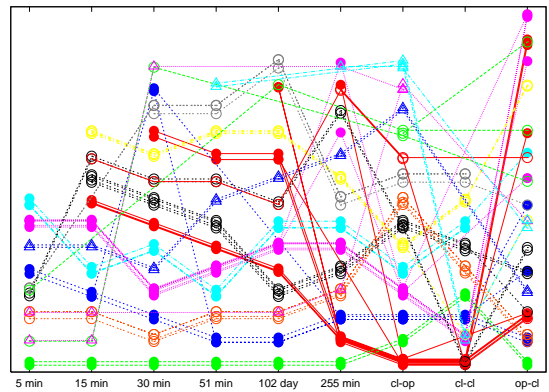} 
      \caption{Evolution of the cluster structure for set $E$ of LSE.}
     \label{evol.cls.e.LSE}
\end{figure}

\begin{figure}[ht]
%  \centerline{\hbox{\psfig{figure=evol.cls.paris.b.all_time.eps,height=6cm,width=8cm}}}
      \includegraphics[scale=0.6] {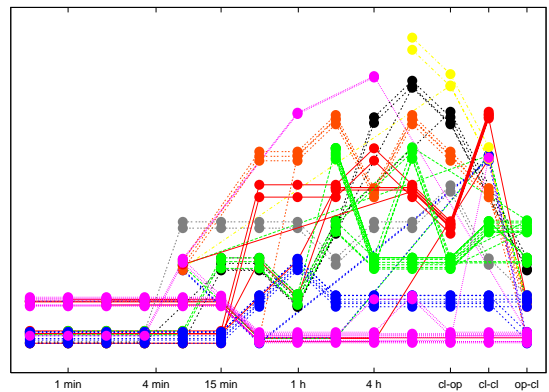} 
      \caption{Evolution of the cluster structure for set $B$ of 75 stocks in PB.}
     \label{evol.cls.e.paris}
\end{figure}

We have performed data clustering analysis also on LSE and PB data. Again we found that removing the market mode allows one to reveal the structure of correlations much more clearly. Indeed, while set $A$ is characterized by one or two clusters at intraday time scales, set $B,\ldots,E$ are characterized by a richer structure, as shown in Figs. \ref{evol.cls.e.LSE} and
 \ref{evol.cls.e.paris}. In both cases, we see that a significant part of the structure forms at intermediate time-horizons of $15$ - $30$ minutes. In PB data, we pushed our analysis to ultra-high frequency, probing very short time scales. We found that for $\tau<$5 min barely any structure can be seen in the correlation matrix. As for NYSE, we found that the cluster structure of set $\hat A$ is poorly correlated with the classification of assets in economic sectors, whereas datasets $B$ and $E$ cluster in a way which reflects up to 70\% of the (entropy of a) classification in economic sectors for LSE, and that this information content is roughly constant across time (intraday) scales. 
 
As for the NYSE, we found that overnight returns have a cluster structure which is markedly different from that of intraday returns. Different markets, however, exhibit different patterns in this respect. While the LSE has a fragmented cluster structure of overnight returns similar to NYSE, PB shows a more compact structure.
 
In contrast with our findings on NYSE data, the cluster structure of set $A$ is now markedly different from that of other sets even at the daily scale.
This suggests that the role of global correlation is much stronger in LSE and PB. 

In order to compare different markets, we performed the Kolmogorov-Smirnov (KS) test  \cite{NumRec} on the distribution of cluster sizes. 
This provides a $p$ value for the hypothesis that two different samples $\{s_i^a\}$ and $\{s_i^b\}$ of cluster sizes can be considered as different populations drawn from the same unknown parent distribution. 
If this is not the case (i.e. $p$ is small), we can conclude that the two samples have a different structure, whereas if $p$ is close to one, we cannot reject the hypothesis that the two samples have the same structure. 
We found that LSE and PB have a cluster size distribution which is different from that of NYSE ($p\simeq 0.1$), but which are remarkably similar one to the other ($p\simeq 1$). 

The similarity between LSE and PB, and their difference with NYSE, is also visible in the dependence of the largest eigenvalues on $\tau$ shown in Fig. \ref{fig:eig}. Remarkably, the market mode seems stronger in NYSE than in LSE and PB, whereas data clustering suggests the opposite.

\begin{figure}[ht]
%  \centerline{\hbox{\psfig{figure=evol.cls.paris.b.all_time.eps,height=6cm,width=8cm}}}
      \includegraphics[scale=0.6] {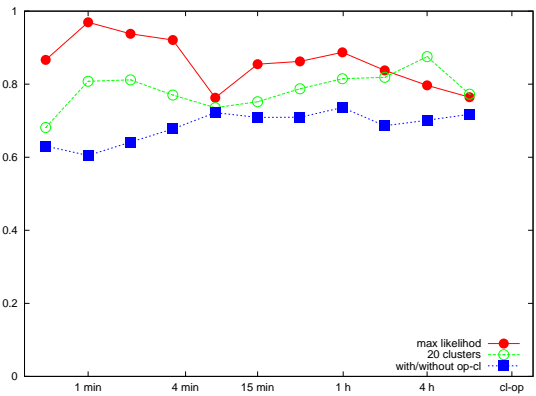} 
      \caption{Relative entropy $\Im$ of the cluster structures of set $B$ of PB obtained for {\em i)} tick and real time (circles) on maximum likelihood structures (filled) or structures with 20 clusters (open) and {\em ii)} with and without the opening and closure period of roughly 30' (filled squares).}
     \label{time_check}
\end{figure}

In the case of PB data, we also performed several tests in order to asses the sensitivity of our results on the inhomogeneity of trading activity. One may indeed think that particular times of the day, such as the opening or the closure of the market, peak or lunch break hours, might be characterized by different statistical properties. In order to test for these effect, we removed the first and the last 20 minutes of trading from the data in each day and considered the resulting correlation matrices $\hat A', \hat B',\ldots$. We computed the relative information $\Im$ between the maximum likelihood structures obtained in this way and the original ones, at different time scales $\tau$. The result is that, for set $B$ of PB, at all $\tau$ roughly $\Im\simeq 70\%$ of the structure found in the whole dataset coincides with that obtained eliminating the opening and the closing period (see Fig. \ref{time_check}). An even stronger similarity ($\Im=0.83$) was found in NYSE between the structure of intraday  correlations and those obtained from returns measured roughly $30$ minutes after opening and before closing. We conclude that a significant part of the structure is not affected by the activity at the market opening or at closure.

As a further test to check the effects of time inhomogeneity of trading activity, we compute correlation matrices in tick time for PB, over intervals of $\tau_k^{\rm (tick)}=100\cdot 2^k$ ticks, which correspond on average to the time scales $\tau_k$ used in real time (here a tick is defined as a transaction on any of the stocks considered). The results, shown in Fig. \ref{time_check}, suggest that the structure of market correlation is largely independent of the definition of time, as indeed roughly $80\%$ of the information found with real time is recovered using tick time. 

%%%%%%%%%%%%%%%%%%%%%%%%%%%%%%%%%%%%%%%%%%%%%%%%%%%%%%%%%%%%%%%%%%%%%%%%%%%%%%%%%%%%%%%%%%%
\section{Single Linkage Clustering Analysis} \label{SLCA}
%%%%%%%%%%%%%%%%%%%%%%%%%%%%%%%%%%%%%%%%%%%%%%%%%%%%%%%%%%%%%%%%%%%%%%%%%%%%%%%%%%%%%%%%%%%

In this section we review the results obtained by applying the Single
Linkage Clustering Algorithm (SLCA) to the data considered in section
\ref{data}. For each time-horizon considered, the SLCA allows to
obtain a Hierarchical Tree (HT) and a Minimum Spanning Tree (MST),
which give complementary information about the network structure of
the considered set of stocks. Indeed, the HT gives a description of
the hierarchical organization of the stocks, while the MST gives an
indication about their topological organization. For a review of SLCA in the context of multivariate financial time series we refer to \cite{AdP,Mantegna1999,EPJB2004}.

As much as in the previous section, here we apply the SLCA to the
different datasets in order to investigate how the structure of
market's correlations evolves as the time-horizon $\tau$ increases
from intraday scales to the daily scale. We shall first focus on NYSE
and then discuss the differences found in other markets. 
The colors used in the representation of both the HTs and the MSTs
refer to the classification is sectors of economic activity given in Table \ref{classification}.

%%%%%%%%%%%%%%%%%%%%%%%%%%%%%%%%%%%%%%%%%%%%%%%%%%%%%%%%%%%%%%%%%%%%%%%%%%%%%%%%%%%%%%%%%%%
\subsection{NYSE}

The investigation of NYSE data by using the SLCA reveals that the role
of the ``center of mass'' in the structure of the correlation is
twofold. On one side, the level of clustering in all the HTs in the
sets where the ``center of mass'' is removed is at an higher distance
than the corresponding HTs of set $A$. Such effect is expected since, by
removing the ``center of mass'', the mean correlation is now
approximately zero, as shown in Fig.  \ref{fig:hcij}. On the other
side, the cluster structure seems now to be more evident than
in the case of the original data. 

\begin{figure}[ht]
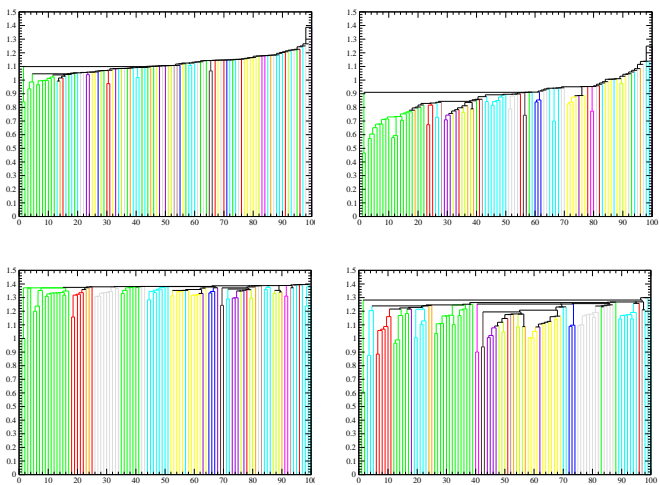

\begin{center}
      \vbox{ 
      \hbox{\includegraphics[scale=0.18]{HT_5.eps}        \hspace{.2 cm}
            \includegraphics[scale=0.18]{HT_intraday.eps}
           }\vspace{.5 cm}
      \hbox{\includegraphics[scale=0.18]{HT_OFM_5.eps}    \hspace{.2 cm}
            \includegraphics[scale=0.18]{HT_OFM_intraday.eps}
           } 
      }
      \caption{HT for set $A$ (top) and $E$ (bottom) of NYSE at $\tau=5$ min (left) and at daily (op-cl) time-horizon (right). The vertical lines represent different stocks. For each stock, colors refer to its economic sector of activity, see table \ref{classification}. Economic sectors of activity are defined according to the classification scheme used in the web--site {\tt{http://finance.yahoo.com/}}.}  
\label{HTNYSE}
\end{center}
\end{figure}

In Figs. \ref{HTNYSE} we present the data for
set $A$ (top) and set $E$ (bottom) at the two extreme time-horizon of 5 min (left) and 1
at 1 day (right). Contrary to what we find in set $A$ (top left), the HT
of set $E$ at 5 min time-horizon (top right) shows a
significant level of structure that, additionally, is
similar to the one found at 1 day (op-cl) time-horizon (bottom right).

\begin{figure}[ht]
\begin{center}
  \includegraphics[scale=0.25]{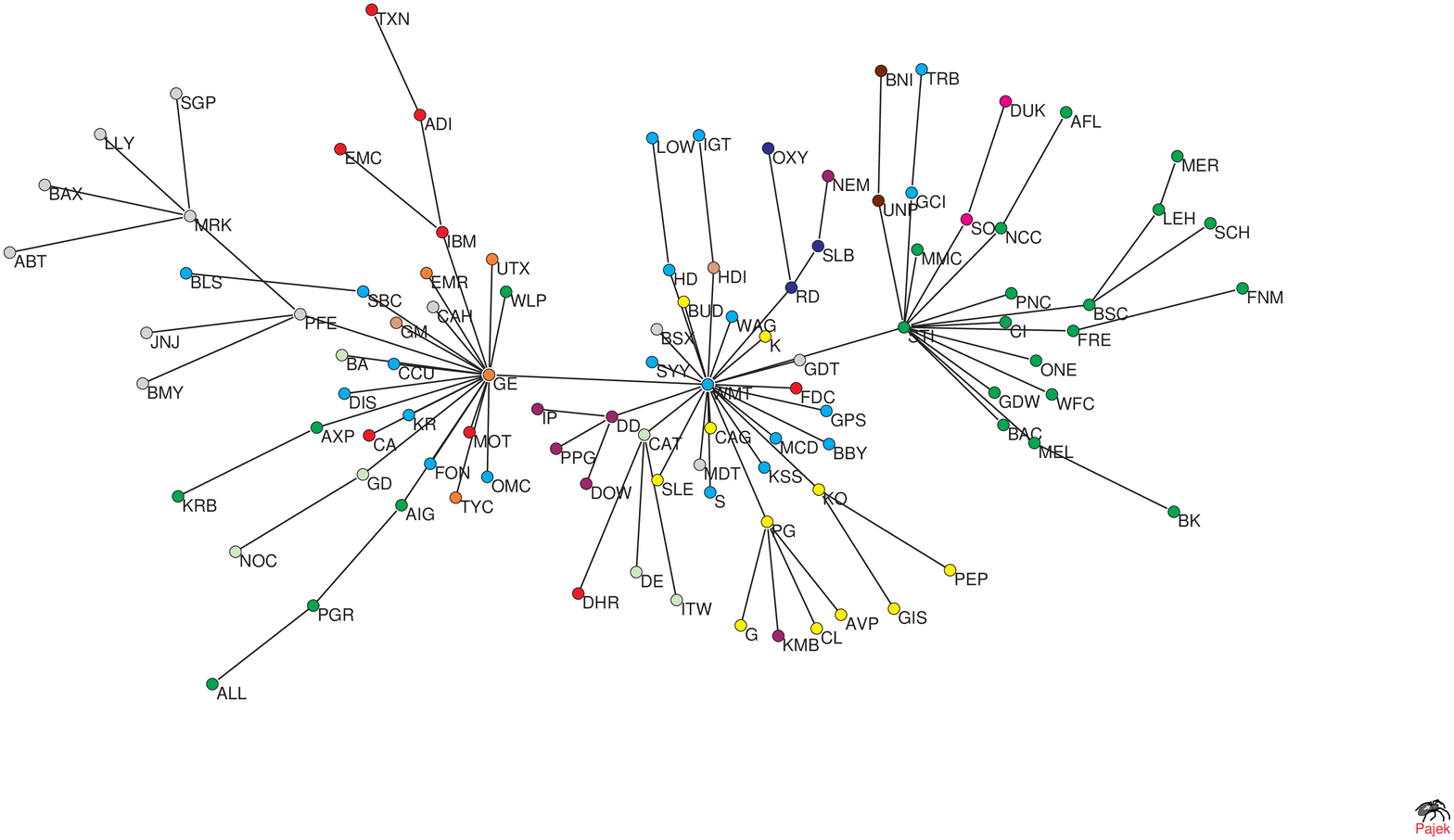}
      \caption{MST  for set $A$ of NYSE at $\tau=5$ min. The vertices represent different stocks. For each stock, colors refer to its economic sector of activity, see table \ref{classification}. Economic sectors of activity are defined according to the classification scheme used in the web--site {\tt{http://finance.yahoo.com/}}.}  
\label{MST5NYSEa}
\end{center}
\end{figure}

\begin{figure}[ht]
\begin{center}
      \includegraphics[scale=0.25]{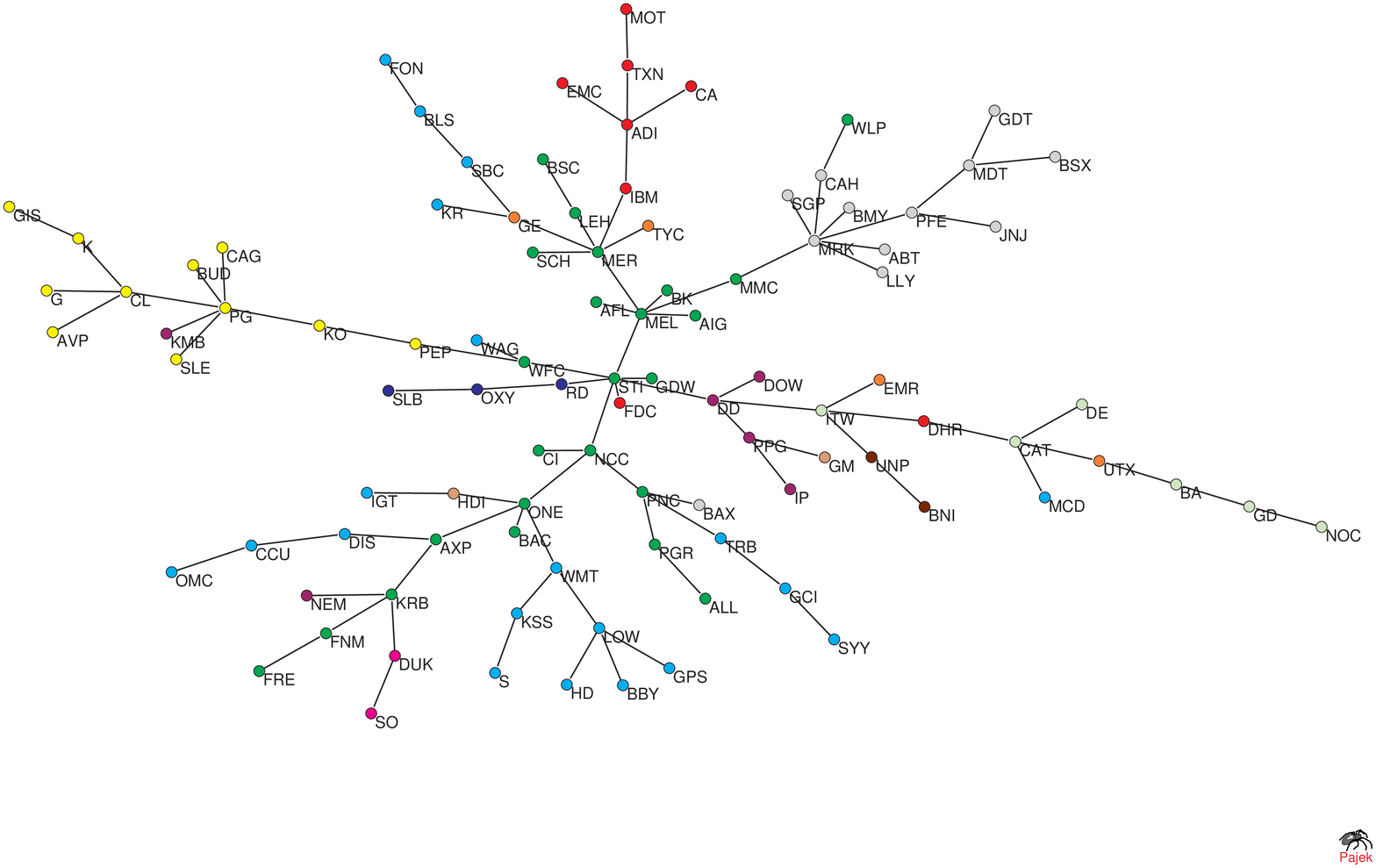}
      \caption{MST  for set $A$ of NYSE at $\tau=$1 day (op-cl). The vertices represent different stocks. For each stock, colors refer to its economic sector of activity, see table \ref{classification}. Economic sectors of activity are defined according to the classification scheme used in the web--site {\tt{http://finance.yahoo.com/}}.}  
\label{MSTdayNYSEa}
\end{center}
\end{figure}

\begin{figure}[ht]
\begin{center}
  \includegraphics[scale=0.25]{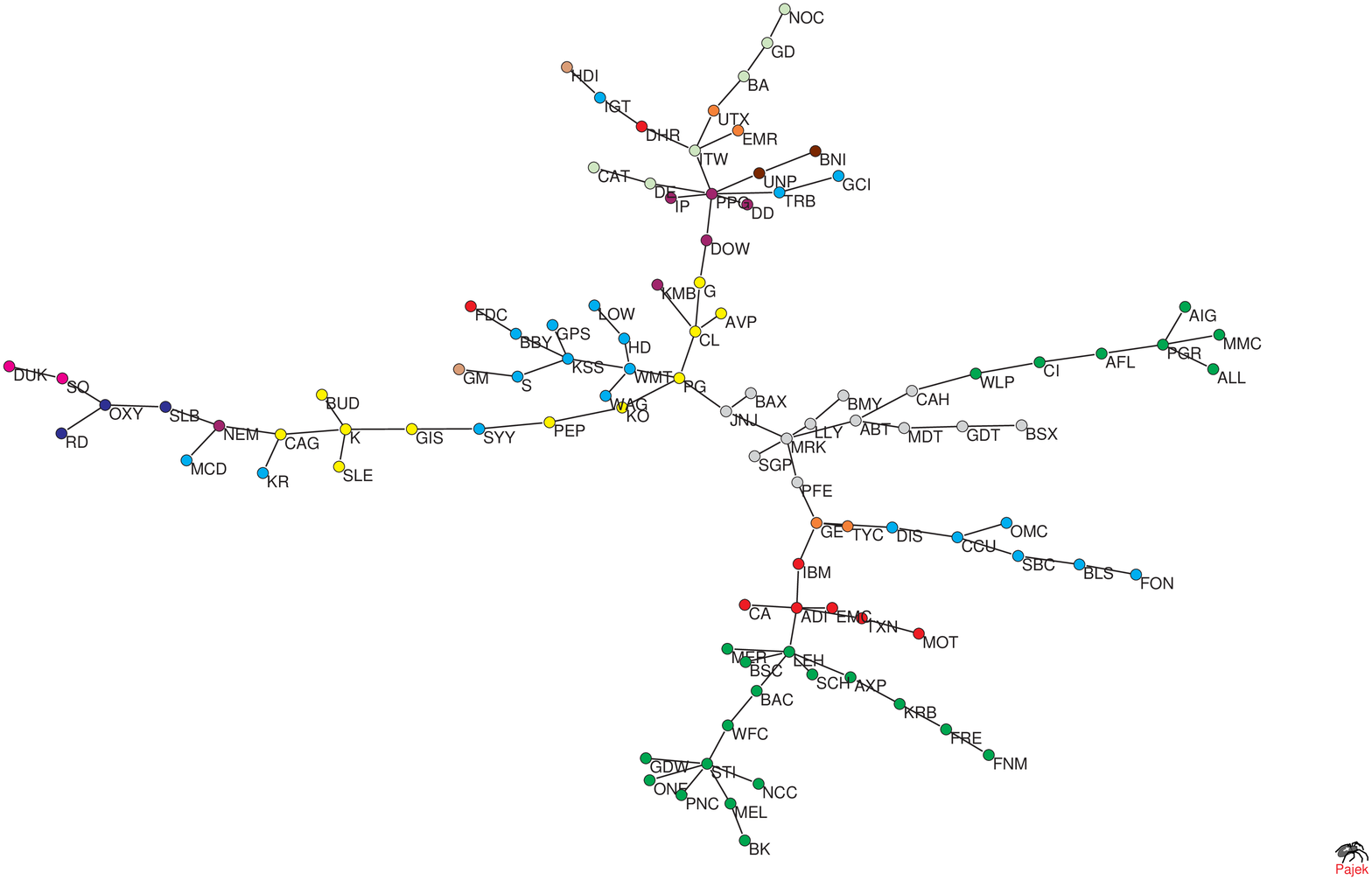}
      \caption{MST  for set $E$ of NYSE at $\tau=5$ min. The vertices represent different stocks. For each stock, colors refer to its economic sector of activity, see table \ref{classification}. Economic sectors of activity are defined according to the classification scheme used in the web--site {\tt{http://finance.yahoo.com/}}.}  
\label{MST5NYSEe}
\end{center}
\end{figure}

\begin{figure}[ht]
\begin{center}
      \includegraphics[scale=0.25]{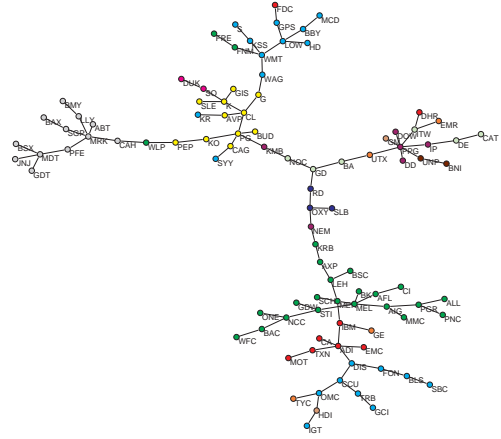}
      \caption{MST  for set $E$ of NYSE at $\tau=$1 day (op-cl). The vertices represent different stocks. For each stock, colors refer to its economic sector of activity, see table \ref{classification}. Economic sectors of activity are defined according to the classification scheme used in the web--site {\tt{http://finance.yahoo.com/}}.}  
\label{MSTdayNYSEe}
\end{center}
\end{figure}

This is also confirmed by comparing the structure of the MST in sets
$A$ and set $E$. The 5 min MST of set $A$ shows a typical structure
with a few hubs characterized by an high degree (Fig. \ref{MST5NYSEa}). The
1 day (op-cl) MST of set $A$ indicates that the number of hubs has
increased, reflecting the progressive organization of stocks according
to their sectors of activity as time-horizon increases (Fig.
\ref{MSTdayNYSEa}). The MSTs shown in Figs. \ref{MST5NYSEe} and
\ref{MSTdayNYSEe} for set $E$ are markedly different from  the corresponding ones for set $A$. 
No preminent hub is traceable in the two MSTs. In addition, they have a structure which is
remarkably similar one another, to the extent that one could not say
which is which, on the basis of their statistical structure alone.

In order to quantify the difference between the structure of the MSTs of different datasets at different time-horizons, we performed the Kolmogorov-Smirnov (KS) test \cite{NumRec} on the degree distributions of MSTs. 
The results for different sets are collected in table \ref{tab:KS} and it largely confirms the conclusions based on visual inspection of Fig. \ref{MST5NYSEa} -- \ref{MSTdayNYSEe}. First, we see that the structure of the MSTs at the extremes of the intraday scale range are markedly different in set $A$ and become increasingly similar as we move to set $E$. Second, Table  \ref{tab:KS} shows that the structure of set $A$ is similar to that of other sets at the same time-horizon at $\tau=1$ day (op-cl), but this is not true at smaller time-horizons. 

We also compared the MSTs with random MST (r-MST) generated by uncorrelated random walks of the same length. This reveals that, apart from set  $A$, we are not able to detect any statistical feature in the degree distribution which differentiates the MSTs of sets $B,C,D$ and $E$ at $\tau=1$ day (op-cl) from those generated by pure noise. Even the diameter of the MSTs is not able to discriminate them from those generated by pure noise. However, the similarity of MSTs with r-MST disappears for larger datasets of $N=500$ or $N=2000$ stocks of NYSE, for which KS yields values of $p\simeq 0$ for all sets, at both $\tau=$5 min and 1 day (op-cl). Furthermore, MSTs turn out to be considerably more compact than r-MSTs. For example, we find that with $N=500$ the r-MST have a diameter of $53$ whereas at $\tau=$1 day (op-cl) the largest value of the diameter is $37$ for set $E$. Finally, in the case of $N=500$ stocks, for set $B$, set $C$, set $D$ and set $E$ we have also performed the KS test in order to compare the degree distribution of the MSTs at 5 min and 1 day (op-cl). Such tests confirm the result of Table $\ref{tab:KS}$, valid for $N=100$ stocks, that the degree distributions are essentially indistinguishable, with p-values which are close to zero. Hence, we conclude that the removal of the ``market mode'' generates residues whose MSTs still contain non-trivial statistical features, although these are not clearly observable in the case of $N=100$ assets. When considering a larger set, say $N=500$, the noise threshold lowers enough to reveal a topological organization which is different from the one associated to uncorrelated random walks.

\begin{table}[htdp]
\begin{center}\begin{tabular}{c|c|c|c|c|c}
~~   & $A_\tau$ & $B_\tau$ & $C_\tau$ & $D_\tau$ & $E_\tau$ \\\hline
$X_{5 min}$, $\tau=1d$ & 0.031 & 0.677 & 0.961 & 0.992 & 1.000\\\hline
$A_{\tau=5 min}$ & 1.000 & 0.047 & 0.021 & 0.001 & 0.000 \\\hline
$A_{\tau=1d}$ & 1.000 & 0.794 & 0.894 & 0.677 & 0.794\\\hline
$R$, ${\tau=5 min}$ & 0.000 & 0.443 & 0.677 & 0.794 & 0.992\\\hline
$R$, ${\tau=1d}$ & 0.556 & 1.000 & 1.000 & 1.000 & 1.000\\\hline
$d$, ${\tau=5 min}$ & 8 & 17 & 16 & 23 & 23\\\hline
$d$, ${\tau=1d}$ & 15 & 26 & 22 & 22 & 25\\\hline
\end{tabular} 
\caption{Results of the Kolmogorov-Smirnov test on the degree distribution of MSTs for different datasets and time-horizons in NYSE. The first row compares the MSTs at $\tau'=$5 min and $\tau=$1 day (op-cl) in different datasets $X=A,\ldots,E$. The second (third) row compares the structure of the MST of set $A$ at $\tau=$5 min (1 day) with the MSTs in different datasets at the same horizon $\tau$. The fourth (fifth) row compares the MSTs of sets $A,\ldots,E$ at $\tau=$5 min (1 day) 
with one generated by a random sample of $N=100$ random walks of the same length. The last two rows report the diameters of the MSTs at $\tau=$5 min and 1 day (op-cl). These should be compared with the diameter $22\pm 3$ of a r-MST generated by uncorrelated random walks.} \label{tab:KS}
\end{center} 
\end{table}

\begin{figure}[ht]
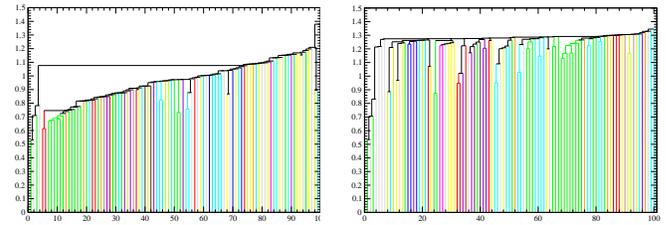

\begin{center}
      \includegraphics[scale=0.18]{HT_overnight.eps} \hspace{.15 cm}
      \includegraphics[scale=0.18]{HT_OFM_overnight.eps}
      \caption{HT for set $A$ (left) and $E$ (right) of NYSE at overnight time-horizon. The vertical lines represent different stocks. For each stock, colors refer to its economic sector of activity, see table \ref{classification}. Economic sectors of activity are defined according to the classification scheme used in the web--site {\tt{http://finance.yahoo.com/}}.}  
\label{overnightNYSE}
\end{center}
\end{figure}

In Fig. $\ref{overnightNYSE}$ we show the HTs relative to set A (left) and set E (right) in the case when the overnight time-horizon is considered. The structure of such trees is different form the ones at intraday time-horizons. In particular, for set A, the HT of Fig. $\ref{overnightNYSE}$ shows that some stocks are highly correlated with each other. However, the organization in economic sectors of activity is less marked than in the corresponding HT at daily time-horizon, see Fig. $\ref{HTNYSE}$. Such effect is also observable when considering set E, i.e. the right panel of Fig. $\ref{overnightNYSE}$. Here the average level of correlation increases, as expected. It is therefore evident that at the overnight time-horizon the organization of stocks in clusters is different than at intraday time-horizons, i.e. when the market is open. 

We have seen above that when removing the market mode the topology of the MSTs has no specific statistical features, even though the distribution of stocks on them is definitely not random. Indeed the cluster structure seen in HTs (Fig. \ref{HTNYSE}) correspond to the fact that companies 
belonging to the same economic sector appear clustered in the same
region of the MST. Again, this shows that the removal of the ``center
of mass'' reveals the organization in sectors of activity already at
such a small time-horizons as 5 min. It is worth remarking, though,
that the location of sectors themselves along the tree is different at
5 min and at the intraday scale. In other words, the
intra-sector structure evolves in time, while the sector composition
remains stable. 

\begin{figure}
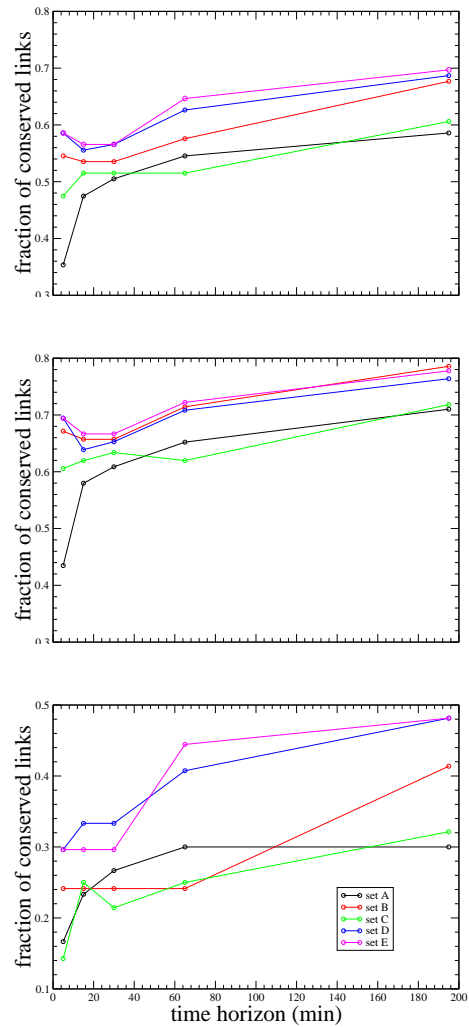

\begin{center}
      {\vbox{
             \includegraphics[scale=0.25]{OVERLAP_all_new.eps}\vspace{.25 cm} 
             \includegraphics[scale=0.25]{OVERLAP_INTRA_new.eps}\vspace{.25 cm} 
             \includegraphics[scale=0.25]{OVERLAP_INTER_new.eps}
      }}
      \caption{Fraction of the intraday MST
links that are conserved with respect to the open-to-close case in the NYSE data. We report the cases where we
consider all the links (top), only intra-sector links (middle) or only
inter-sector links (bottom). Economic sectors of activity are defined according to the classification scheme used in the web--site {\tt{http://finance.yahoo.com/}}.}  \label{overlap}
\end{center}
\end{figure}

In order to give a quantitative description of this effect, for each
set and at each time-horizon we have measured the fraction of the MST
links that are conserved with respect to the open-to-close case. The
results are reported in Fig. \ref{overlap}. The top panel refers to the case when all links in the MST are considered.
The other two panels refer to the case when we also use the information about the economic sectors of activity, see Table \ref{classification}. In particular, we consider only intra-sector links (middle panel) or only
inter-sector links (bottom panel). Ideally, for a better quantitative description we should have considered clusters rather than economic sectors. Unfortunately, the SLCA does not allow a precise identification of what a cluster is. However, in Fig. $\ref{fig:relent}$ it is shown that there exists a strict relation between economic sectors and the clusters obtained by using the methodology of Ref. \cite{giada}. We here somehow make the ansatz that such strict relation persists also in the clusterization given by the SLCA.

In the case when we consider all links
(top) or only those between stocks in the same sector (middle), in all
the cases but one, when the center of mass has been removed, the
fraction of conserved links is higher than for set $A$.  The middle
panel of Fig. \ref{overlap} shows that 70-80\% of the MST links
between stocks belonging to the same economic sector are conserved
with respect to the open-to-close case, whereas a much smaller
fraction is conserved between stocks belonging to different economic
sectors. This is consistent with our observation that while sector
composition remains stable, intra-sector correlations evolve with the
time-horizon. Moreover, such results are also consistent with the ones shown in Fig. $\ref{fig:relent}$ that the amount of economic information contained in the clusters is constant.

In this respect, the botton panel of Fig. \ref{overlap} shows that set $D$ and set 
$E$ reveal better than the others the topogical organization within
different economic sectors at all time-horizons. Finally, it is worth remarking that 
set $C$, where the market mode is {\em{exogenously}} given by the SP500 index, gives results which are comparable with those of set $A$.

By summarizing, the investigation of sets $A$, $B$, $C$, $D$ and
$E$ by using the SLCA shows that (i) the removal of the ``center of
mass'' reveals the organization of the sectors within different
economic sectors even at small time-horizons and (ii) this is better
achieved in set $D$ and set $E$, where the ``center of mass'' is
{\em{endogeneously}} obtained either by miminizing the $\chi^2$ function of
Eq. \ref{endo} or by using a mere return market average. Finally, we find that the degree distributions of the MST at different time-horizons are
statistically the same, specially  in set $E$, according the the Kolmogorov-Smirnov test, but they cannot be distinguished from those of a set of $N$ independent random walks, for such a small market ($N=100$). The distribution of stocks on the MST reflects the organization of stocks in economic sectors, and indeed links between companies in the same sector are ``conserved'' across time scales. 

%As a general comment, it must be emphasized that our results confirm how the information associated to the clusters or economic sectors of activity is not beared by the mean value of the cross--correlations $\langle \rho_{ij}\rangle$, i.e. the first moment in the distribution of cross-correlation coefficients. In fact, as shown in Fig. $\ref{fig:hcij}$, the pdf of cross-correlation coefficients for set B, set C, set D and set E are centered around the null value and still those sets give rise to non-trivial MSTs and HTs. On the other hand, by comparing the above four sets of data with a set of $N$ independent random walks for which $\langle \rho_{ij}\rangle=0$, it is evident that the relevant information is beared by the higher moments $\langle \rho_{ij}^{~q} \rangle$ of the distribution of cross-correlation coefficients. %%% SALVATORE: SCUSA SE HO COMMENTATO QUESTO PARAGRAFO. E' CHIARO CHE LA DISTRIBUZIONE (MARGINALE) DELLE Cij NON DA' NESSUNA INFORMAZIONE SULLA STRUTTURA. LA TRUTTURA, PIUTTOSTO CHE NEI MOMENTI PIU' ALTI DELLA PDF DEI Cij STA NELLA PDF DEGLI ELEMENTI DI MATRICE DELLE POTENZE DELLA MATRICE C 

%%%%%%%%%%%%%%%%%%%%%%%%%%%%%%%%%%%%%%%%%%%%%%%%%%%%%%%%%%%%%%%%%%%%%%%%%%%%%%%%%%%%%%%%%%%

\subsection{Other markets}

The question arises whether the above results have some degree of universality or they are peculiar to the NYSE market. We have therefore repeated the above investigations for different markets, i.e. for LSE, PB and BI. Generally we confirmed the main conclusions: We find that HT of sets $B,C,D$ and $E$ reveal better the organization of stocks in economic sectors than set $A$, and that the structure of HTs for the formers is less dependent on the time-horizon $\tau$ than for the latter. 
The structure of MSTs has a clear evolution in set $A$ as the time-horizon increases (e.g. KS test yields $p_{LSE}=0.051$ for the degree distributions of MSTs of set $A$ between $\tau=$5 min and 1 day), whereas it has a remarkably stable structure in the other sets (particularly for set $E$, for which $p_{LSE}=0.999$ between $\tau=$5 min and 1 day). A comparison of the MST for set $E$ for NYSE and LSE yields a KS test value of $p>0.9$ for all time-horizons $\tau$. Similar results were found comparing NYSE and PB or BI MSTs. This invariance of the structure of MSTs for set $E$ across markets and time-horizons should not be considered as an indication of universality, though. Indeed, as for NYSE, this invariant structure is indistinguishable from that of r-MSTs generated from uncorrelated random walks. 
Hence, what this allows us to conclude is that markets of such small sizes do not allow to make statements on the similarity of market topology in terms of their MSTs. Indeed, the topology of MSTs for $N\approx 100$ stocks or less, is dominated by noise.

\begin{figure}
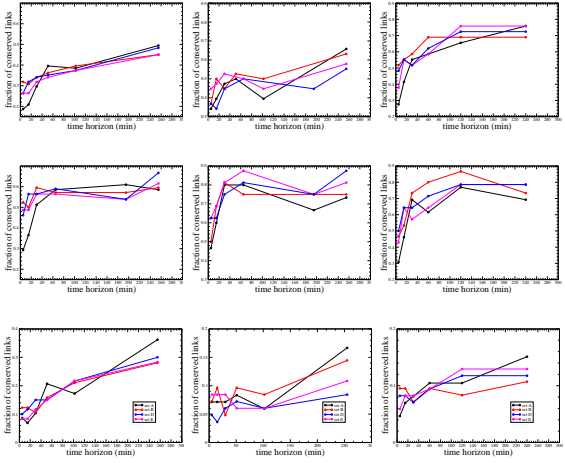

\begin{center}
      {\vbox{
             \includegraphics[scale=0.10]{OVERLAP_all_new_LSE.eps}\vspace{.2 cm} 
             \includegraphics[scale=0.10]{OVERLAP_all_new_CAC.eps}\vspace{.2 cm} 
             \includegraphics[scale=0.10]{OVERLAP_all_new_MIB.eps}\\
             \includegraphics[scale=0.10]{OVERLAP_INTRA_new_LSE.eps}\vspace{.2 cm} 
             \includegraphics[scale=0.10]{OVERLAP_INTRA_new_CAC.eps}\vspace{.2 cm}
             \includegraphics[scale=0.10]{OVERLAP_INTRA_new_MIB.eps}\\ 
             \includegraphics[scale=0.10]{OVERLAP_INTER_new_LSE.eps}\vspace{.2 cm}
             \includegraphics[scale=0.10]{OVERLAP_INTER_new_CAC.eps}\vspace{.2 cm}
             \includegraphics[scale=0.10]{OVERLAP_INTER_new_MIB.eps}             
      }}
      \caption{Fraction of the intraday MST
links that are conserved with respect to the open-to-close case in the LSE (left), PB (middle) and BI (right) data. We report the cases where we
consider all the links (top), only intra-sector links (middle) or only
inter-sector links (bottom). Economic sectors of activity are defined according to the classification scheme used in the web--site {\tt{http://www.euroland.com/}}.}  \label{overlapother}
\end{center}
\end{figure}

When the market is open, the disposition of stocks on the MSTs, as in NYSE, is consistent with economic classification, across time-horizons. In Figs. \ref{overlapother} we report, for different sets and at each time-horizon, the fraction of the MST links that are conserved with respect to the open-to-close case for LSE (left) and PB (middle) and BI (right). Again, we consider all the links (top), only intra-sector links (middle) or only inter-sector links (bottom). As much as in the NYSE case, the sectors considered here are the economic sectors of activity mentioned above. In the case of LSE data the results are less sharp than in the NYSE case. Set $B$, set $D$ and set $E$ give results which are more similar to each other with respect to the NYSE case. One possible exception is given by set $B$ at 5 min time-horizon. In all cases, it is confirmed that the removal of the ``market mode'' reveals the organization of stocks in economic sector already at small time-horizons. As an example, the fraction of conserved intra-sector links in set $E$ is always ranging between $50 \%$ and $60\%$, while in set $A$ such percentage drops to $30 \%$ at the smallest time-horizon. At larger time scales. however, the fraction of conserved links for set $A$ has roughly the same value that for other sets. This is different from what we found for NYSE, where the fraction of conserved links were systematically smaller for set $A$ than for other sets. 

When considering the overnight time-horizon, we confirm that the organization of stocks in economic sectors of activity is less evident than in the case when the market is open. However, such differences are less marked than in the NYSE case.

%%%%%%%%%%%%%%%%%%%%%%%%%%%%%%%%%%%%%%%%%%%%%%%%%%%%%%%%%%%%%%%%%%%%%%%%%%%%%%%%%%%%%%%%%%%
\section{Conclusions}
%%%%%%%%%%%%%%%%%%%%%%%%%%%%%%%%%%%%%%%%%%%%%%%%%%%%%%%%%%%%%%%%%%%%%%%%%%%%%%%%%%%%%%%%%%%

We found that removing the dynamics of the center of
mass {\em i)} decreases the level of correlations and {\em ii)} makes
the cluster structure more evident. Na\"ively one would expect that
reducing the level of correlations reduces the ``signal'' and hence
enhances the role of noise in the dataset. On this ground, one might
expect a less sharply defined structure, i.e. the opposite of {\em
  ii)}. The fact that we observe {\em i)} and {\em ii)} implies that
the market mode dynamics bears little or no information on the market
structure. It also suggests that the market mode dynamics and the
dynamics of ``internal coordinates'' are to a large extent separable,
in much the same manner as in particle systems of classical mechanics, where the center of mass dynamics accounts for the effect of external forces, whereas relative coordinates respond to internal forces arising from inter-particle potentials.

It is not difficult to imagine components of trading activity which might contribute to the dynamics of the ``center of mass'' or to relative coordinates. It is worth to remark, in this respect, that a simple phenomenological model for the dynamics of the market mode, taking into account the impact of trading in risk minimization strategies, has been recently proposed \cite{pfolio}. Besides reproducing the main statistical properties of the dynamics of the largest eigenvalue of the covariance matrix, this model also shows that the behavior of the market mode is largely insensitive to a finer structure of correlations. The invariance of the structure of  ``internal'' correlations across time scales, and its similarity with economic classification, instead suggests that the dynamics of relative coordinates might be related to the ways in which information on different assets diffuses in the market.

The finding of a scale-invariant correlation structure is non-trivial, in several respects. First, its origin suggests a fine balance between signal and noise across time-horizons:
On one hand, the growth of correlations implicit in the Epps effect implies that the ``signal'' gets stronger as the time scale increases. On the other, random matrix theory suggests that the strength of ``noise'' due to finite sampling, is more severe at large time-horizons than at short ones. Indeed, the length of the time series decreases as $T\sim 1/\tau$, which implies a spread $\delta\lambda\sim\sqrt{N/T}\sim\sqrt{\tau}$  in the eigenvalues due to noise dressing. This latter effect allows us to detect weak correlation structures with an high precision at small time scales. 

Secondly, the scale invariance of correlation structure might have important implications for risk management, because it suggests that correlations on short time scales might be used as a proxy for correlations on longer time-horizons. 
If the structure of correlations at short time scales can be computed using shorter time series, this might allows us to detect structural changes more efficiently.

Finally, uncovering the dynamical origin of such a complex phenomenology poses exciting challenges to theoretical modeling of multi-asset markets. 

%%%%%%%%%%%%%%%%%%%%%%%%%%%%%%%%%%%%%%%%%%%%%%%%%%%%%%%%%%%%%%%%%%%%%%%%%%%%%%%%%%%%%%%%%%%%%%%%%%%%%%%%%%
\section{Acknowledgments}
Authors acknowledge support from research projects MIUR 449/97 ``High frequency dynamics in financial markets", M.M. acknowledges support from EU-STREP project n. 516446 COMPLEXMARKETS. S. M. acknowledges support from MIUR-FIRB RBNE01CW3M ``Cellular Self-Organizing nets and chaotic nonlinear dynamics to model and control complex systems", from the EU-STREP projects n. 012911 ``Human behavior through dynamics of complex social networks: an interdisciplinary approach". We wish to thank Dr. Claudia Coronnello for assistance in the preparation of data.

%%%%%%%%%%%%%%%%%%%%%%%%%%%%%%%%%%%%%%%%%%%%%%%%%%%%%%%%%%%%%%%%%%%%%%%%%%%%%%%%%%%%%%%%%%%%%%
%%%%%%%%%%%%%%%%%%%%%%%%%%%%%%%%%%%%%%%%%%%%%%%%%%%%%%%%%%%%%%%%%%%%%%%%%%%%%%%%%%%%%%%%%%%%%%

\end{document}